\documentclass[11pt,twoside]{article}
\usepackage{CAGN2019}
\usepackage{graphicx}
\usepackage[T1]{fontenc} % Computer Modern (CM) fonts
\usepackage{latexsym}
\usepackage{verbatim}
\usepackage{ifpdf}
\usepackage{amsmath,amssymb}
\ifpdf
      \DeclareGraphicsExtensions{.pdf,.png,.jpg}
\else
      \DeclareGraphicsExtensions{.eps}
\fi
\newcommand\ion[2]{#1$\;${\scriptsize\rmfamily{#2}}\relax}

\newcommand{\msun}{M$_\odot$}

\begin{document}

\vskip 1.0cm
\markboth{L.~Carigi et al.}{Galactic Disk Models based on H II Region abundances}
\pagestyle{myheadings}
\vspace*{0.5cm}
\parindent 0pt{Contributed  Paper}

\vspace*{0.5cm}
\title{
Chemical evolution models for the Galactic disk based on \ion{H}{II}
region abundances
derived from a direct method and a temperature independent method
}

\author{L.~Carigi$^1$, M.~Peimbert$^1$, and A.~Peimbert~$^1$}
\affil{
$^1$Instituto  de Astronom\'\i a, Universidad Nacional Aut\'onoma de M\'exico, Apdo. Postal 70-264, M\'exico, C.P. 04510, CdMx, Mexico\\
}

\begin{abstract}

We present two chemical evolution models of our galaxy, both models are built to fit the O/H ratios derived from
\ion{H}{II}  regions, using two different methods.
One model is based on abundances obtained from the [O III] 4363/5007 temperatures (direct method, DM) and
the other on abundances obtained from the recombination line ratios of [O II/H I] (temperature independent method, TIM).
The differences between the O/H values obtained from these two methods are about 0.25~dex.
We find that the model based on the TIM values produces an excellent
fit to the observational stellar constraints (B-stars, Cepheids, and the Sun),
while  the model based on the DM fails to reproduce each of them.
Moreover, the TIM model can explain the flattening of the O/H gradient observed in the inner disk
due to the assumption of an inside-out star formation quenching, in the $3-6$ kpc galactocentric range, starting $\sim9$ Gyr ago.


\end{abstract}

%------------------------
\section{Introduction}

A chemical evolution model (CEM) computes the chemical abundances of the gas mass in a Galactic zone
and the predicted evolution can be tested by comparing it with the chemical abundances of no polluted stars which ages are known.
The inferred chemical history of the Galactic zone will be more reliable,
when the observational constraints  become more precise.

In this work,
we present two CEMs for the Galactic disk obtained using the CHEVOL code that considers ingredients
dependent on time, $t$(Gyr), galactocentric distances, $R$(kpc), and metallicity of the gas, $Z$:
such as galactic flows, star formation laws, initial mass functions, and stellar properties of each formed star.
The code considers the lifetime of each star, independently of it mass.

For more details of these CEMs and their observational constraints, see \cite{Carigi2019} (CPP19).

%------------------------
\section{Ingredients of the Chemical Evolution Models }

The main characteristics of our CEMs are:
i) The Galaxy is assembled with primordial infalls but without any type of outflows.
During the first Gyr the halo formed efficiently, then the disk formed with $R$-dependent efficiencies following an inside-outside scenario. The amount of accreted gas for each $R$ is chosen to reproduce $Mtot(R)$.
ii) The star formation rate is the well known law $ {\it SFR}(R,t) \propto Mgas^{1.4}(R, t)$ by
%\cite{kennicutt1998}.
\cite{Kennicutt2012}.
The proportionality coefficient is a spatial and temporal function, $ \nu (Mgas +Mstar)^{0.4}$,
and $\nu$ is obtained to fit the general behavior of $Mgas(R)$ and the flattening of the O/H gradient for $R < 6$ kpc.
iii) The initial mass function is that by \cite{Kroupa1993}
in the $0.08 - Mup$ \msun  \ range.  $Mup$ is a free parameter of the models,
that is obtained to match the absolute value of the preset-day O/H gradient.
iv) Each star enriches the interstellar medium when it leaves the main sequence,
depending of its initial stellar mass and metallicity. Moreover, SNIa are taken into account.
v) The age of the models is 13 Gyr and the Sun is located in an orbit of 8-kpc-medium radius.
vi) Radial flows of gas or stars are not included.

%------------------------
\section{Observational data to built the CEMs}

The models are built to reproduce the following observational radial distributions along the Galactic disk:
i) the total baryonic mass, $Mtot(R)$, ii) the gas mass, $Mgas(R)$, and iii) the O/H values from \ion{H}{II}  regions.
For $Mtot(R)$, we adopt an exponential profile with a disk scale length  $ \sim$ 4 kpc, and for $Mgas(R)$, we take into account the distribution by \cite{Kennicutt2012}. Regarding the O/H($R$), we consider O/H gaseous values determined from the direct method (DM) and the temperature independent method (TIM),
and we include the fraction of O atoms trapped in dust grains.

For the DM, we consider the O/H gaseous values for 21 \ion{H}{II}  regions derived from the forbidden lines (FLs) of O$^{++}$ under the assumption that there are no temperature inhomogeneities; see \cite{Esteban2018} and Table 1 by CCP19.
The dust contribution used depends slightly on $Z$ and is in the $0.10 - 0.11$ dex range \citep[][]{Peimbert2010}.

For the TIM,  the O/H values are determined from the recombination lines (RLs) of O$^{++}$ and H$^+$ (10   \ion{H}{II} regions, see Table 2 by CCP19) and we used the \cite{Peimbert2010} dust correction.
For objects where no RLs are available (11 \ion{H}{II}  regions, see Table 3 by CCP19) we correct (O/H)$_ {FL}$ values due to temperature inhomogeneities and presence of dust by adopting the calibration by \cite{Penaguerrero2012}.

\cite{  }

\begin{figure}  %%%%%%%FIGURE 1 %%%%%%
\begin{center}
\hspace{0.25cm}
\includegraphics[height=18.0cm]{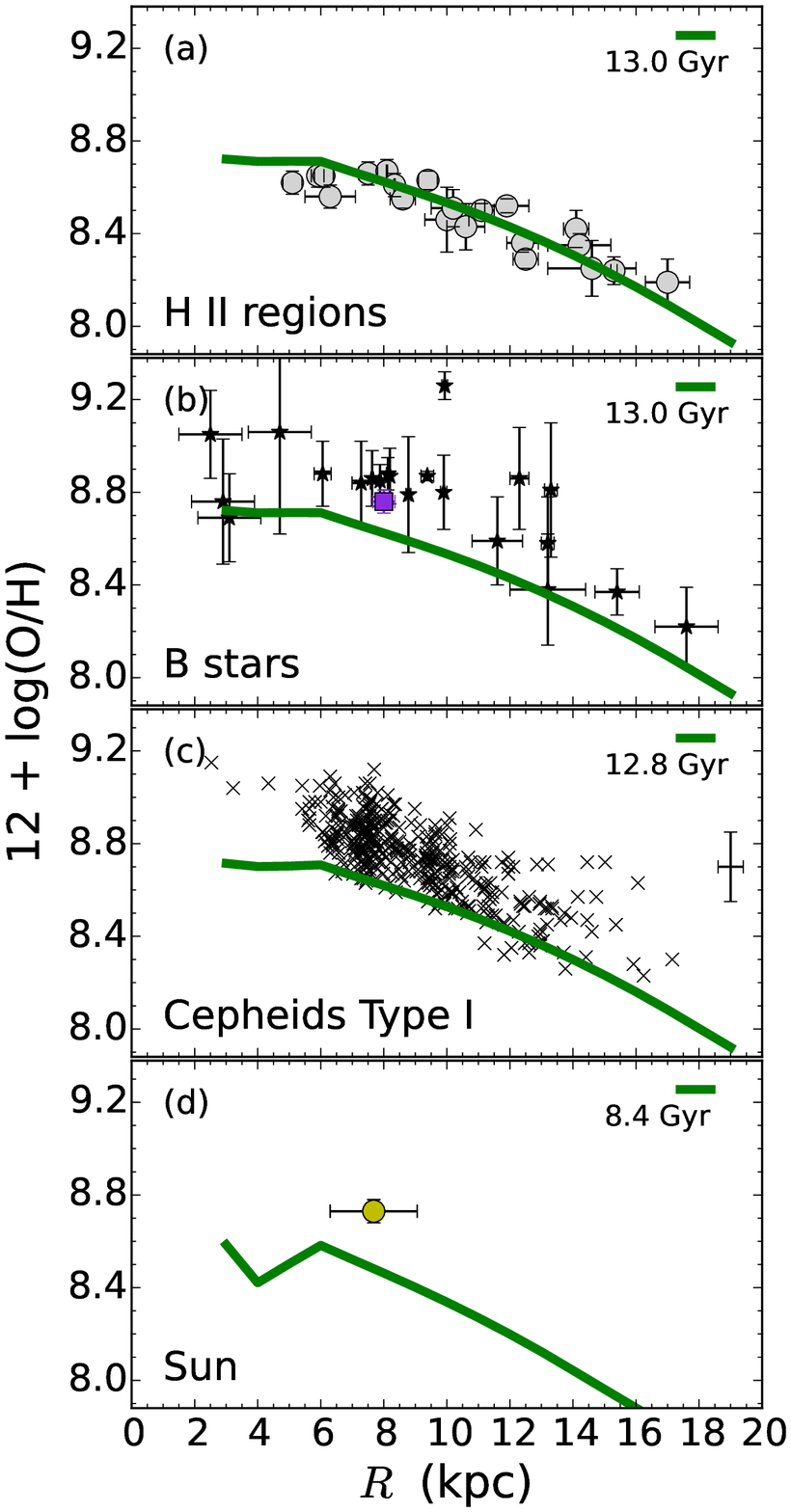}
\caption{
Radial distribution of O/H obtained by the DM model at different evolutionary times (green lines).
Panel (a): H II regions. Gaseous values obtained with the Direct Method plus dust corrections.
Panel (b):  B-stars. Black stars: values by \cite{Rolleston2000}  and \cite{Smartt2001}. Violet square:  the average value by \cite{Nieva2012}.
Panel (c): Cepheids of Type I. Error bars: uncertainties for the typical Cepheid.
Panel (d): Initial solar value by \cite{Asplund2009}. The horizontal error bar: average migration of the Sun computed by \cite{Martinez2017}.
}
\label{DM}
\end{center}
\end{figure} %%%%%%%FIGURE 1 %%%%%%

%------------------------
\section{Observational data to test the CEMs}

Early B type stars are very young objects, with ages less than $\sim$ 100 Myr
and their abundances can be considered as good representatives of the present-day abundances. In this work, we consider the mean O/H value of 13 B-stars of the Orion OB 1 association by \citet{Nieva2012},
located at distances from the Sun smaller than 350 pc.
Moreover, we take into account a set of 18 O/H values in the $6-18$ kpc range obtained from 51
early type B-stars of Galactic open cluster associations by \citet{Rolleston2000}.
We also present 4 B-stars in the $2-5$ kpc range, studied by  \cite{Smartt2001}.

Cepheids of Type I are younger than 200 Myr, and
consequently their O/H values can be compared with the predicted O/H values at 12.8 Gyr.
We considered a set of 397 disk Cepheids located in the $3-17$ kpc
range compiled by Martin (private communication, 2017).

The age of the Sun is $\sim$ 4.6 Gyr and the protosolar abundance
\citep[12 + log(O/H) = 8.73 dex,][]{Asplund2009}
would represent the O/H of the ISM at 8.4 Gyr. However, its birth galactocentric distance could be different than the current one, because the Sun might have migrated. In this work, we take into account the dynamic models by \cite{Martinez2017} that indicate that the Sun may have been born in the $6.3 - 9.1$ kpc range.

%------------------------
\section{The Direct-Method model}

The DM model is built to fit the O/H values of the \ion{H}{II}  regions derived from the direct method.
These values are reproduced by considering  $Mup=40$ \msun.
Moreover, the model requires that $\nu=0.019$ to match the general behavior of $Mgas(R)$,
however with this constant $\nu$  this model predicts an O/H gradient steeper than observed for $R < 6$ kpc. To reproduce the O/H flattening for $R < 6$ kpc,
we assume a unit step function, such that  diminishes abruptly the {\it SFR(R)} at specific quenching times as a function of  $R$.
Then we adopt the current star formation rate observed at these radii
\citep{Kennicutt2012}  for the rest of the evolution.
Specifically, to reproduce the 12+log(O/H) $\sim 8.70$ value for $R = $3, 4, and 5 kpc, the quenching times are 3.30, 3.50, and 5.40 Gyr, and the present-day $SFR$ is equal to 0.45, 0.55, and 0.65 \msun pc$^{-2 }$Gyr$^{-1}$, respectively.
We ran several models, testing different quenching times, where we chose the highest times that match the $ 8.70 $ dex value.

In Fig.1 we present the results of this model built to fit the O/H values of the \ion{H}{II}  regions from the DM (panel a) and we test it with O/H stellar abundances (panels b, c, and d).
In Fig.1(b, c), we compare the predicted value of the O/H($R$) for the present time and for 0.2 Gyr ago with the values from B-stars and Cepheids, respectively.
It can be noted that the agreement is poor, being the predicted values $\sim 0.25$ dex lower than observed.
In Fig.1(d), we present the predicted O/H($R$) at 8.4 Gyr,  the protosolar O/H value,
and the initial solar orbit.
The predicted initial solar value is in the 8.40 to 8.56 range, about 0.25 dex smaller than the inferred one. No stellar migration (outward or inward) can explain the O/H solar value.

\begin{figure}  %%%%%%FIGURE 2%%%%%%
\begin{center}
\hspace{0.25cm}
\includegraphics[height=18.0cm]{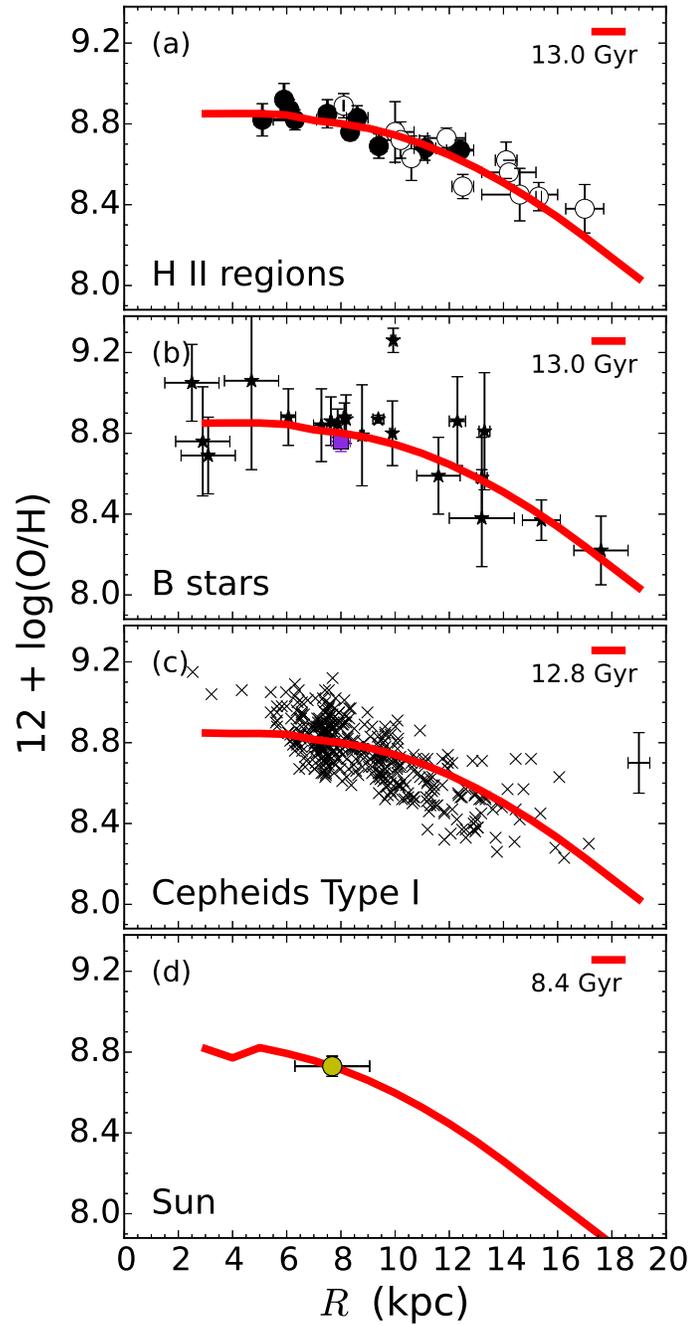}
\caption{
Radial distribution of O/H obtained by the TIM model at different evolutionary times (red lines).
Panel (a): H II regions. Gaseous values obtained with the Temperature Independent Method plus dust corrections. The filled circles denote direct observations of the recombination lines, the empty circles use the calibration by
\cite{Penaguerrero2012}.
Panels (b), (c), and (d) as Fig. 1.}
\label{TIM}
\end{center}
\end{figure}  %%%%%%FIGURE 2%%%%%%

%------------------------
\section{The Temperature-Independent-Method model}

The TIM model is built to reproduce the observed O/H values
based on the temperature independent method.
Since the (O/H)$_{TIM}$ values are higher  by $\sim 0.25$ dex than (O/H)$_{DM}$,
the TIM model needs more massive stars to produce O, consequently an
$Mup=80$ \msun \ is required.
We apply the same idea and procedure of quenching that in the DM model,
because the $R$ behavior of the O/H values from both methods are very similar.
To reproduce the 12+log(O/H) $\sim 8.83$ value for $R = $3, 4, and 5 kpc, the quenching times are 4.08, 6.10, and 8.38 Gyr,  respectively; later times than those inferred by the DM model, because the O/H flattening value from the TIM is higher.

In Fig.2(a) we present the model built to fit the O/H values of \ion{H}{II}  regions
derived from the TIM.
In Fig.2(b) we compare the predicted current O/H($R$) with O/H values in B-stars at different $R$ and we find that the model is in good agreement with these observations.
In Fig.2(c) we test the predicted O/H($R$) at 0.2 Gyr ago with the Cepheid data.
The match is very good for $R > 6$ kpc, but for the inner disk more data are needed.
In Fig.2(d) we present the predicted O/H radial distribution at 8.4 Gyr, the protosolar value
and the possible $R$s where the Sun was born. The spatial, temporal, and O/H agreements with the model are excellent, and consequently, the TIM model does not need to invoke a radial migration of the Sun to explain its chemical properties.

%------------------------
\section{Conclusions}

We present two galactic chemical evolution models:
i) the Direct Method model (DM model), built to fit the O/H values in H II regions derived from the
forbidden lines of O$^{++}$, dependent on temperature;
and ii) the temperature-independent-method model (TIM model),  built to fit the O/H values based on the recombination lines of O$^{++}$and H$^+$. Moreover corrections due to O embedded in dust grains are considered.  The (O/H)$_{DM}$ values are smaller by $\sim 0.25$ dex than (O/H)$_{TIM}$.

The DM model cannot match either the early B-stars and the Cepheid data, or the solar O/H value,
even considering an average migration for the Sun.
In particular, the predicted O/H values are typically smaller, by $\sim $  0.25 dex, than the stellar chemical abundances.

The TIM model reproduces very well the O/H values from B  and  Cepheid stars. Moreover,
the predicted O/H when the Sun was formed is in excellent agreement with the protosolar O/H value,
without having to invoke stellar migration.

If the O/H flattening in the $3-6$ kpc range is corroborated,
it would imply an inside-out star formation rate quenching in this range, that started $\sim$ 9 Gyr ago.

\acknowledgments L. C. thanks the support from CONACyT, grant 247132.

\bibliographystyle{aaabib}
\bibliography{bib}

\end{document}